\documentclass[prb,twocolumn,showpacs]{revtex4}
\usepackage{graphicx}
\usepackage{dcolumn}
\usepackage{bm}

\begin{document}
\title{Low frequency excitations of C$_{60}$ chains inserted inside single-walled carbon nanotubes}
 
\author{J. Cambedouzou, S. Rols, R. Almairac, J.-L. Sauvajol}
\affiliation{Groupe de Dynamique des Phases Condens\'ees (UMR CNRS 5581), Universit\'e Montpellier II, F-34095 Montpellier Cedex 5, France.}%

\author{H. Kataura}
\affiliation{Nanotechnology Research Institute, National Institute of Advanced Industrial Science and Technology Central 4, Higashi 1-1-1, Tsukuba, Ibaraki 305-8562, Japan.}

\author{H. Schober}
\affiliation{Institut Laue Langevin, F-38042 Grenoble, France}

\date{\today}
\newcommand{\degree}{$^{\circ}$}%

\begin{abstract}

The low frequency excitations of C$_{60}$ chains inserted inside single-walled carbon nanotubes (SWNTs) have been studied by inelastic neutron scattering (INS) on a high quality sample of peapods. The comparison of the neutron-derived generalized phonon density of states (GDOS) of the peapods sample with that of a raw SWNTs allows the vibrational properties of the C$_{60}$ chains encapsulated in the hollow core of the SWNTs to be probed. Lattice dynamical models are used to calculate the GDOS of chains of monomers, dimers and polymers inserted into SWNTs, which are compared to the experimental data. The presence of strong interactions between C$_{60}$ cages inside the nanotube is clearly demonstrated by an excess of mode density in the frequency range around 10 meV. However, the presence of a quasi-elastic signal indicates that some of the C$_{60}$'s  undergo rotational motion. This suggests that peapods are made from a mixture of C$_{60}$  monomers and C$_{60}$  {\it n}-mer (dimer, trimer ... polymer) structures.

\end{abstract}
\pacs{61.12.-q, 61.46.+w, 61.48.+c, 63.22.+m, 78.67.Ch}
\maketitle
The insertion of  C$_{60}$ in single-walled carbon nanotubes (SWNTs) has recently attracted considerable attention because of the predicted superconducting properties upon alkaline metal doping\cite{Saito}. Electron microscopy images \cite{Smith} of these so-called 'peapods' have revealed the one-dimensional character of the C$_{60}$  chains inside the carbon nanotubes. Raman investigations have been performed \cite{KatauraA}, and it was found that a photo-polymerization of the C$_{60}$ under a 488 nm laser excitation wavelength occurs. The latter observation suggests that C$_{60}$ are spinning inside the hollow core of the tubes at room temperature. Raman spectra have therefore been recorded at helium temperature to derive the peapods response before polymerization. The spectra were found to be stable, indicating that no photopolymerization occured {\it during} the measurements. However, a mode located at $\sim$ 90 cm$^{-1}$ was observed at low temperature, the frequency of which is close to the dumbell-like mode of a C$_{60}$ dimer (96 cm $^{-1}$). X-Ray \cite{KatauraA,nous?} (XRD) and electron \cite{Hirahara,KatauraA} (ED) diffraction measurements at room temperature have stated that the inter-C$_{60}$ distance ranges from 9.5 to 9.8 \AA\ in highly filled samples. This distance is consistent with van der Waals interactions between fullerenes as indicated by recent calculations \cite{Rochefort}. However, the XRD derived thermal expansion of the C$_{60}$ chains was found to be much smaller than the one of the 3D-crystalline phase \cite{KatauraA}, and comparable to the in-plane dilation for graphite. All these results show that the precise nature of the C$_{60}$-C$_{60}$ bond in peapods is ambiguous: pure structural data suggests a van der Waals (VDW) character while methods involving dynamical properties (thermal expansion, vibrations) reveal a much stronger, polymer-like bond nature. This ambiguity led the C$_{60}$  chains to be described as a {\it loosely bonded polymer structure}. In this letter, we present the results of an INS study performed on a peapods powder. The results will be discussed based on previous INS investigations on both SWNTs \cite{Rolsprl} and polymeric C$_{60}$ \cite{Schober}. In addition, we present lattice dynamics calculations of the GDOS of different possible structures for encapsulated C$_{60}$ chains.\\
%
A 900 mg peapods sample was specially prepared for the INS experiment using the sublimation method reported elsewhere \cite{KatauraSM}. The characterization of this sample was performed by X-Ray and neutron diffraction on a small part and on the totality of the sample, respectively. No differences between the patterns were found which indicates a good homogeneity of the sample. The powder is found to contain rather large bundles of at least 20-30 tubes with a very narrow diameter distribution centered around 13.5 \AA\ , a filling rate of roughly 80\% and a 9.8 \AA\ inter-C$_{60}$ distance \cite{nous?}. A 300 mg raw SWNT sample used as reference for this INS study was also synthesized by the arc-discharge method. The XRD derived structural parameters of this sample are similar to those of the peapods sample {\it e.g.} the bundle parameters (tube diameter, number of tubes and diameter distribution) can be considered to be equivalent for both samples.\\

%
Inelastic neutron scattering is the ideal experimental technique for measuring the frequency range where inter-C$_{60}$ interactions dominate the spectrum. This technique has been succesfully used to study the dynamics of several phases of pure and alkaline-doped C$_{60}$ \cite{Schober,Kolesnikov}. They allowed lattice dynamical models to be developed and force constants to be refined \cite{Schober}. In particular, it was found that the inter-molecular modes region (below 8 meV) of the 3D-crystalline state of C$_{60}$ GDOS is clearly separated from the intra-molecular modes region (above 30 meV) by a $\sim$22 meV gap. The dimer and polymer phases of C$_{60}$ obtained upon alkaline doping or under high temperature and high pressure conditions are characterized by a progressive filling of the gap region of the GDOS\cite{Schober,Kolesnikov}. This region provides therefore a useful tool to probe the C$_{60}$ bonding scheme.\\
The peapod sample and the reference nanotube sample have been studied using the IN6 Time Of Flight (TOF) spectrometer at the Institut Laue Langevin (Grenoble) which has high flux and good resolution for measuring the GDOS in the [0-50 meV] energy range. A neutron wavelength of 4.12 \AA\ was used and the measurements were performed at 300 K and 480 K giving reasonable intensity in the neutron energy gain (anti-Stokes) spectrum. Raw TOF data were treated using the usual procedure to obtain the $S(\vec{Q},\omega)$ scattering law. From this quantity, both the imaginary part of the generalized susceptibility $\omega^{-1}\chi''(\omega)$ and the generalized density of states  $G(\omega)$  can be derived \cite{Schober}. Both quantities are known to be appropriate functions when studying dynamically induced disorder and vibrations. Finally, the GDOS of the SWNTs was subtracted from that of the peapods after a proper normalization was performed, in order to get the GDOS of C$_{60}$ ch
 ains inserted inside the SWNTs. The normalization accounts for incomplete tube filling and difference in total vibrational mode number from both species \cite{explication}. \\
%
%
Figure 1 shows $\omega^{-1}\chi''(\omega)$ measured at 300\,K and 480\,K for the peapod and the nanotube samples. To obtain reasonable statistics the data have been summed over all angles. For harmonic phonons, the curves should overlap at all temperatures for each sample, as Debye-Waller and multi-phonon corrections compensate each other. The nanotube sample clearly shows such harmonic behaviour. In contrast $\omega^{-1}\chi''(\omega)$ for the peapod sample at 480\,K is inferior to $\omega^{-1}\chi''(\omega)$ at 300\,K. This relative excess of intensity at low temperature is a typical signature of relaxation motion giving rise to a quasi-elastic signal that doesn't follow the bose statistics. Phonon softening due to anharmonicity would show the opposite trend and thus cannot account for the observed temperature evolution. The assignment of part of the low-frequency spectrum to relaxation motion in the peapod sample is further reinforced by the general shape of the signal. It shows the typical form observed for C$_{60}$ in the plastic phase if determined under identical experimental conditions. The complete study of this quasi-elastic contribution (as Q, and T dependence) cannot be performed due to its weak intensity and the limited $Q$-range. However, isotropically rotating  C$_{60}$ molecules or short n-mers (dimers or trimers) rotating about the tube axis are likely to be responsible for this quasi-elastic signal.\\ 
\begin{figure}[h!]
\includegraphics[scale=0.7]{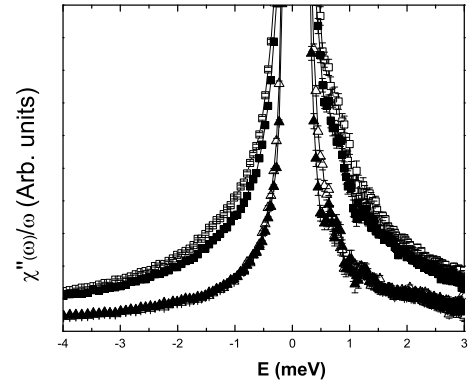}
\caption{\label{fig:figure2}  Susceptibility of the peapods (squares) and of the nanotubes samples (triangles) measured at 300 K (open symbols) and 480 K (filled symbols).}
\end{figure}
%
%
The normalized GDOS of the peapods, SWNTs and inserted C$_{60}$ at 480 K are shown in Figure 2a. The GDOS of inserted C$_{60}$ does not go below zero at any frequency and this confirms the hypothesis of a weak coupling between the vibrational modes of the tube and those of the C$_{60}$   in the low energy range. Recent lattice dynamics calculations \cite{Damn} have shown that the coupling between the radial breathing modes (RBM) of the two constitutive tubes of a (5,5)@(10,10) double-walled carbon nanotube (DWNT) is weak. According to the similar transverse dimensions of a peapod and a (5,5)@(10,10) DWNT, the non-observation of any coupling in the peapods spectrum is therefore not surprising. The general shape of the GDOS of the inserted C$_{60}$ chains displays two regions separated by a gap around 20 meV. The low energy range is reported on the top of Figure 3. The GDOS in this range features a peak at 4 meV, a shoulder at $\sim$ 1.8 meV and a large unresolved band around 10 meV. Concerning the energy range above the gap, a rather intense peak is clearly observed around 33 meV which is characteristic of the H$_{g}$(1) intramolecular mode of the C$_{60}$ balls.\newline
\begin{figure}[h!]
\includegraphics[scale=0.55]{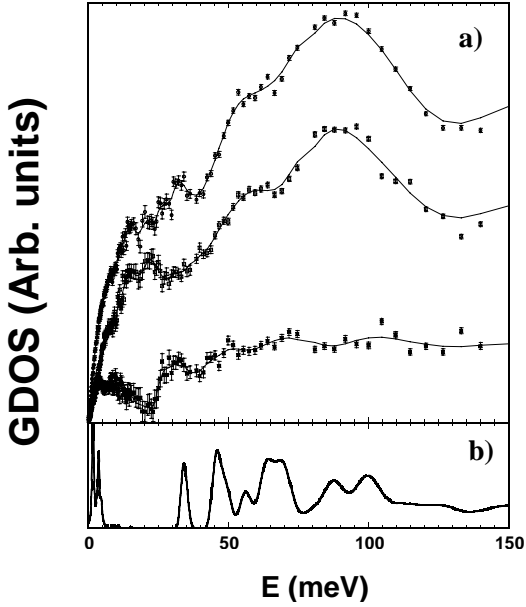}
\caption{\label{fig:figure2}  a) Generalized density of states (GDOS) of the peapod sample (top curve),  of the reference SWNT sample (middle curve) and of inserted C$_{60}$ chains (bottom curve) resulting from the substraction of the GDOS of SWNTs from the GDOS of peapods after adequate normalization. All measurements were performed at 480 K. Lines are guides to the eyes resulting in a three points smoothing of the data. b) Calculated GDOS of a chain of C$_{60}$ monomers inserted into a SWNT. This line is the result of the substraction of the GDOS calculated for empty nanotubes from the GDOS calculated for the whole peapods.}
\end{figure}
%
%
The calculations of the GDOS were performed assuming a force field model for both intratube \cite{Saito2,Rolsprl} and intra-C$_{60}$ \cite{Jishi} interactions. Three C$_{60}$ packings were considered in this study: a monomer chain (M model), a dimer chain (D model) and a polymer chain (P model). All calculations were performed at 0 K. In particular, free rotations are frozen at this temperature for the monomer model. In the case of the dimer and of the polymer chains, a [2+2] cycloaddition covalent bonding scheme between the carbon cages was assumed, as described in Schober {\it et al.} models \cite{Schober}. It consists in the opening of two double-bonds belonging to adjacent, oriented C$_{60}$'s and the shifting of the 4 carbon atoms participating in the cycloaddition giving two C-C interfullerene single bonds of 1.58 \AA\  and two C-C intrafullerene single bonds of 1.56 \AA\ . The force constant values used to describe the bonds in the neighbourhood of the cycloaddition are given in Table II of reference 15. The van der Waals (VDW) interactions between a C$_{60}$ and the tube, and between adjacent C$_{60}$   in the same tube  (for the monomer model exclusively) were taken into account using a Lennard-Jones potential, U(R)=C$_{12}$/R$^{12}$-C$_{6}$/R$^{6}$, with parameters C$_{12}$=22500 eV \AA$^{12}$ and C$_{6}$=15.4 eV \AA$^{6}$. These parameter values have recently been found to correctly describe the interaction between C$_{60}$ and graphene, and to reproduce the VDW contribution to the C$_{60}$ bulk cohesive energy reasonably well \cite{Ulbricht}. Each numerical GDOS was convoluted with the energy-transfer dependent resolution function of the IN6 time-of-flight spectrometer and the same procedure as the one used in the experimental data treatment was used to calculate the GDOS of the whole peapod and derive the GDOS of the inserted C$_{60}$ chain. The resulting GDOS calculated for these three structures are shown in Figure 3 and compared to the experimental GDOS obtained in the same energy range. \newline
%
%
\begin{figure}[h!]
\includegraphics[scale=0.45,angle=90]{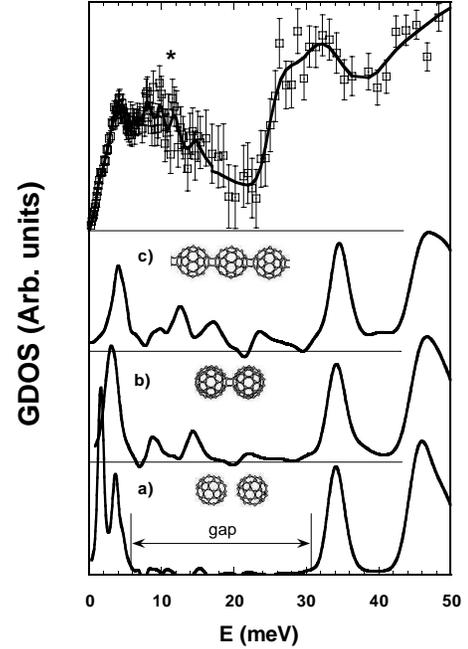}
\caption{\label{fig:figure3}  Lattice dynamical calculations of the GDOS of chains of a) monomers (M model), b) dimers (D model) and c) polymers (P model). The open squares are the experimental GDOS of the inserted C$_{60}$ chain. The line on the experimental spectrum is a guide to the eyes. The star indicates the position of the mode observed by Raman spectroscopy at low temperature.}
\end{figure}
The profile of the GDOS calculated for the M model is displayed in Figure 2b in an enlarged scale. A large gap from 6 meV to 30 meV separates the intermolecular modes below 6 meV from the intramolecular modes above 30 meV. The lowest energy intramolecular mode H$_{g}$(1) is furthermore clearly visible around 33 meV. In the low energy range (see Figure 3a), the GDOS profile consists in two peaks located at 1.7 meV and 3.5 meV. The first peak is related to the C$_{60}$ librations while the second peak involves translational vibrations of C$_{60}$. The comparison between the experimental GDOS and that calculated for the M model shows many differences. First, the absence of the sharp librational peak at 1.7 meV can be understood if C$_{60}$ monomers undergo free rotations. In such a case no libration modes are expected. Second, the gap in the calculated spectrum is too large, especially at the lower energies where additional contributions are clearly observed in the region from 8 to 14 meV. This lack of agreement leads us to consider other structures for the C$_{60}$ chains.\newline
The GDOS calculated for the D model shows important modifications compared to the M model. In the energy range below 6 meV, the two sharp peaks in the monomer spectrum are replaced by a broad peak centered around 3 meV. In the D model, the cylindrical geometry of the C$_{120}$ dimers implies that the librational modes in transverse and longitudinal directions in the tube environment are no longer equivalent so the sharp structure of the libration peak is lost. The frequencies of the translation modes are shifted to slightly lower energy due to the larger mass of the C$_{120}$ dimer. Consequently, librations and translations of the dimer are included into the unique large peak centered around 3 meV. In the region between 8 and 30 meV, two additional peaks located around 9 and 14 meV appear. According to our calculations, these peaks are attributed to the Einstein dumbbell mode of the C$_{120}$ dimer and to libration-like modes where the balls undergo small amplitude rotations perpendicular to the dimer axis, respectively. When the GDOS of the polymer structure is considered, a hardening of the first peak, which shifts to 4 meV, and the increase of the intensities of the modes between 8 and 20 meV relative to the first peak are observed. For this totally polymerized state, the libration of the balls around the chain axis is harder than in the dimer state and contributes to fill the gap. Also, the dumbbell like mode becomes a translational mode having a strong dispersion along the tube axis, giving rise to a broad component in the GDOS in the gap region. \newline
%
%
From Figure 3, it is obvious that none of the models perfectly matches the experimental GDOS. This is not surprising given the simplicity of the models and the fact that the measurements were performed at high temperature. A more sophisticated model should include the diffusive (rotations and translations) motions of the balls as well as disorder effects. However, several important conclusions can be drawn from the comparison between the data and the models. A chain formed by C$_{60}$ monomers in VDW's interaction does not have vibrational modes in the 6-15 meV energy range. This is at variance with the experiment. The GDOS calculated for dimer and for polymer chains demonstrate that  inter-fullerene modes in polymeric structures fill the 6-15 meV energy range. This suggests that at least some of the C$_{60}$ balls are covalently bonded. Long C$_{60}$ polymers in the nanotube would not be in agreement with the average inter-fullerene distance determined by diffraction (9.8 \AA\ $\gg$ 9.2 \AA) and the presence of a quasi-elastic  signal. However a mixed model of monomers and dimers would be compatible with the 9.6 \AA\ inter-fullerene distance. In particular, the modes in the experimental GDOS involved in the large feature around 11 meV  include the inter-C$_{60}$ Raman active stretching bond mode that is calculated at $\sim$9.5 meV in our model and which has been observed at  $\sim$ 11 meV \cite{KatauraA} (located by the star in Figure 3). Since no radiation-induced polymerization can be achieved using neutrons, our observations indicate that dimer-like structures are formed inside the nanotubes during the synthesis.\newline
In summary, the GDOS of C$_{60}$ chains inserted inside SWNT has been extracted from the measurement of the GDOS of a peapod sample and of an empty SWNT sample by inelastic neutron scattering. The experimental GDOS is discussed in the light of lattice dynamics calculations performed on three different models for the C$_{60}$ organization inside the tubes. In particular, intense modes on the low-energy side of the gap separating the intermolecular from the intramolecular modes suggest the existence of strong bonds, such as valence interactions between part of the monomers. The latter observation and the presence of a dynamical disorder induced quasi-elastic signal lead us to propose that peapods are filled with a mix of monomer and {\it n}-mer (dimer, trimer ... polymer) structures, the relative proportion of which is still to be determined. \\
H.K. acknowledges for a support by Industrial Technology Research Grant Program in '03 from New Energy and Industrial Technology Development Organization (NEDO) of Japan. The authors acknowledge Dr. M. Johnson for his help in writing this article.

\vskip 7mm

\newpage


\begin{thebibliography}{99}
\bibitem{explication} The unit cell of a peapods consists of four (10,10) tube unit cells (160 atoms) and one C$_{60}$. This leads to consider that 60/220 of the intensity comes from the C$_{60}$ chain ($\sim$ 30\%). When a 80\% filling of the tubes is assumed, this proportion falls to $\sim$ 20\%.
\bibitem{explication2} The cold neutron source used by IN6 limits the measurements to rather small scattering vector $\parallel \vec Q\parallel $, for which the dynamic structure factor for C$_{60}$ isotropic rotation is weak
\end{thebibliography}
\end{document}